# Automated Detection of Rest Disruptions in Critically Ill Patients

Vasundhra Iyengar, Azra Bihorac, Parisa Rashidi

*Abstract*— Sleep has been shown to be an indispensable and important component of patients' recovery process. Nonetheless, the sleep quality of patients in the Intensive Care Unit (ICU) is often low, due to factors such as noise, pain, and frequent nursing care activities. Frequent sleep disruptions by the medical staff and/or visitors at certain times might lead to disruption of the patient's sleep-wake cycle and can also impact the severity of pain. Examining the association between sleep quality and frequent visitation has been difficult, due to the lack of automated methods for visitation detection. In this study, we recruited 38 patients to automatically assess visitation frequency from captured video frames. We used the DensePose R-CNN (ResNet-101) model to calculate the number of people in the room in a video frame. We examined when patients are interrupted the most, and we examined the association between frequent disruptions and patient outcomes on pain and length of stay.

*Clinical Relevance*— This study shows that rest disruptions can be automatically detected in the ICU, and such information can be used to better understand the sleep quality of patients in the ICU.

## I. INTRODUCTION

Patients in the ICU experience a disrupted circadian rhythm due to factors such as noise, pain, and frequent nursing care activities, thereby impacting their quality of sleep [1]-[3]. The most evident circadian rhythm in humans is the sleep-wake cycle [4]. A disturbed sleep-wake cycle can lead to a lack of adequate rest, thus delaying the patient's recovery time.

The most common disruptions in terms of noise disturbance faced by patients in the ICU are conversations of nurses/other staff, alarms, and medical team interventions [5]. Disruptions by the medical team while unavoidable for the most part, could also have an impact on the patient's sleep cycle and severity of pain. The timing of medical interventions should be analyzed such that their results can be maximized, without disturbing the patient adversely. Hospital visitation policies seldom take these factors into account and are usually based on the staff's presumptions and institutional practices [6]. Previous work for automatic detection of the number of visits is based on small sample sizes (e.g. three patients, [7]), or do not consider the relationship between patient outcomes and frequent visitations [8][9].

In this study, we examined the association between disruptions represented by the number of people present in the room and other factors like pain score and length of stay in the ICU.

## II. METHODS

We automatically detected the number of people present in the room using computer vision deep learning models and we correlated the results with clinical data obtained from electronic health records (EHR) to examine the association between disruptions and patient outcomes.

### A. Data Collection

We conducted an Institutional Review Board (IRB) approved clinical study at the University of Florida Shands Hospital, Gainesville (IRB 201400546). Prior written consent was obtained from the patients enrolled in the study. The study excluded minors (under 18 years) and those with the inability to provide written informed consent. Patients enrolled were admitted to the surgical or medical ICU at the Shands Hospital. A recording system with Amcrest ProHD 1080p POE video cameras with a wide 90° viewing angle was placed in the patient room such that it captured the entire ICU room (Fig. 1). We implemented a comprehensive privacy policy to ensure privacy protection, including posting signs indicating "recording in session", on-demand deletion mechanism, privacy cover for the camera, and strict no-internet access on a strongly encrypted recording computer. We developed a custom user-friendly recording tool to record videos. The recording tool had features to start, stop, and pause the recordings, such that the recording could be stopped whenever needed. Each video segment was recorded for a duration of 15 minutes. Video data were recorded for 38 patients throughout their stay in the ICU, up to 7 days.

Pain Scores were assessed by the ICU nurses as part of the clinical routine on a scale of 0-10 with 0 as no pain, 1-4 as mild pain, 5-6 as moderate pain and 7-10 a severe pain, according to the Defense & Veterans Pain Rating Scale (DVPRS) scale [10]. Each pain score was accompanied by the timestamp denoting the time of pain assessment. We have assumed that a pain score is valid for at least the 20 minutes preceding the pain score entry. Sleep quality was assessed using the Freedman Sleep questionnaire [11]. Age, gender, type of surgery, ICU room details were obtained from EHR.

This work is supported by NSF CAREER 1750192 (PR), and NIH/NIBIB 1R21EB027344 (PR, AB).

V. Iyengar is with the University of Florida, Gainesville, FL 32611 USA (email: vasundhraiyengar@ufl.edu)

A. Bihorac is with the University of Florida, Gainesville, FL 32611 USA (email: abihorac@medicine.ufl.edu)

P. Rashidi is with the University of Florida, Gainesville, FL 32611 USA (corresponding author, phone: 352-392-9469; email: parisa.rashidi@ufl.edu)

## B. Data Sampling

The videos recorded were all timestamped. Videos recorded between 7 am and 7 pm were analyzed for disruption detection, to ensure enough lighting and robust results. Further, the frames were down-sampled at a rate of 1 fps to increase processing speed, as there is no significant change between consecutive frames within a second. For pain score correlations, videos 20 minutes prior to pain assessment timestamps were considered for the association analysis.

TABLE I. PATIENT CHARACTERISTICS.

| Patient Characteristics | All Participants |
|---|---|
| Age, Median (years) | 62 |
| Female, number (%) | 11 (28.9%) |
| Male, number (%) | 27 (71.1%) |
| ICU Length of Stay, Median days | 13.3 |
| Sleep Quality, Median[a] | 4.5 |

a. Calculated for 30 patients

## C. Disruption Detection

For disruption detection, we used the DensePose R-CNN model [15]. It uses the architecture of Mask-RCNN [12] along with the Feature Pyramid Network (FPN) [13] features and Regions of Interest (ROI) Align pooling which provide labels of dense parts and their coordinates within the selected regions. There is a fully convolutional network on top of the ROI-pooling which performs two tasks: For every surface part, generation of per-pixel classification, and regression of local coordinates within each part. The DensePose model was run on all the sampled frames and the number of people detected was extracted. The pre-trained model used in this study was DensePose ResNet-101 FPN trained on DensePose-COCO, a large-scale ground-truth dataset made with 50,000 manually annotated COCO images [14] for dense image to surface correspondences. The inference from DensePose (ResNet-101) pre-trained model ran at 1.5 fps for 1680 x 1050 and 1920 x 1080 images on NVIDIA GeForce Titan X (Pascal) GPU. Evaluation of DensePose-RCNN performance on COCO minival subset for DensePose (ResNet-101) with key-points provides an Average Precision of 87.5 and Average Recall of 91.1 [15]. The end-to-end pipeline for disruption detection is demonstrated in Fig. 2. A sample inference image after running DensePose R-CNN is shown in Fig. 3.

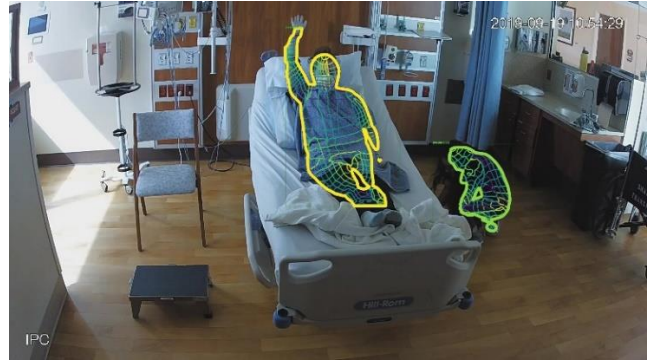

Fig. 3. Sample image after running inference using DensePose ResNet-101 FPN. Actors depicted.

## III. RESULTS

The DensePose R-CNN system detected the number of people in each sampled frame. The total number of frames evaluated for the videos of 38 patients were 5,039,353. Out of these 925,655 frames were evaluated to have zero-person detection and 2,200,709 frames had more than one person detected, defined as a rest disruption of the patient. A total of 43.69% of frames consisted of disruptions. The average number of people in the room when being disrupted was found to be 2.6. The standard deviation for the disruption data in frames was calculated to be 0.87.

### A. Pain Score vs Average Percentage of interruptions

The pain score assessments were correlated with the average percentage of disruptions prior to pain assessments. For each of the patient videos corresponding to a pain score, the number of disruptions was calculated for the preceding 20

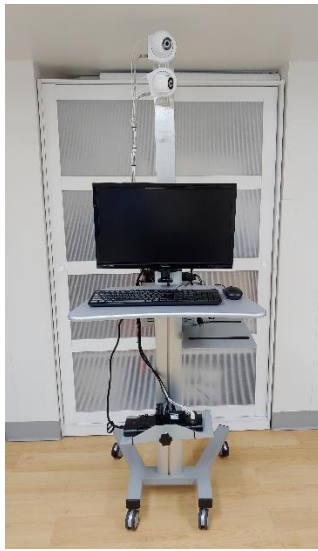
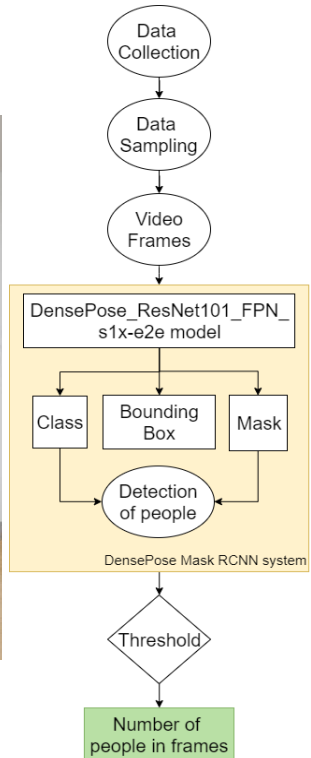

1)            2)

Fig. 1. The recording system used for recording videos in the ICU.

Fig. 2. End-to-end pipeline for disruption detection

minutes. The average number of disruptions versus assessed pain scores can be seen in Fig. 4.

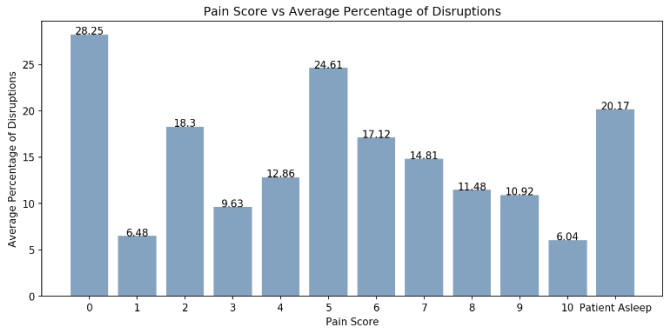

Fig. 4. Pain Score vs Average Percentage of disruptions. Pain Score of 0 represents no pain.

*B. ICU Length of Stay vs Percentage of Disruptions*

The length of stay was calculated for each patient based on admission time to each room the patient was in. A gap of fewer than 24 hours in a non-ICU room was counted as ICU time provided the patient was in the ICU before and after that. These were plotted against the percentage of disruptions averaged for each patient in Fig. 5. We used LOESS smoothing to obtain a smooth best-fit line of the scatter plot to enable relationship analysis. Linear regression of ICU length of stay as the dependent variable and percentage of disruptions was done by including factors such as quality of sleep, gender, age, and type of surgery. We obtained a negative correlation coefficient of -0.43 for the percentage of disruptions and a corresponding P-value of 0.039.

*ICU type Distribution.* The room type a patient is in is classified as Acute, Intermediate, and Intensive depending on the severity of patient care required. The duration of time spent in each care type is averaged for all the patients in the study and shown in Fig. 6.

*Percentage of disruptions per patient.* The percentage of disruptions for each patient was calculated with respect to the total number of sampled frames. For each patient, the percentage can be seen in Fig. 7.

*Percentage of disruptions at different time intervals.* The room type a patient is in is classified as Acute, Intermediate, and Intensive depending on the severity of patient care required. The total duration spent in the ICU was distributed into these classifications and the amount of time a patient spent in each of them can be seen in Fig. 8.

## IV. DISCUSSION

On average 20.17% of disruptions occur when the patient is asleep, as can be seen in Fig. 4. These disruptions can disturb the patient's sleep-wake cycle. Adequate rest is essential for the patient's early recovery and fewer disruptions would possibly improve a patient's recovery time. Maximum disruptions occurred when the patient was in no pain (Pain Score=0) which is the ideal approach. However, the difference between the disruptions for no pain and patient asleep states is only 8.08%. Ideally, fewer disruptions should occur when the patient is asleep. During severe pain (Pain Score=10), the least disruptions occur which implies that the disruptions were likely related to medical care. During mild pain, 24.61% of disruptions occurred. In our future studies, we plan to investigate to understand the impact of disruptions during mild pain in relation to the patient's recovery time.

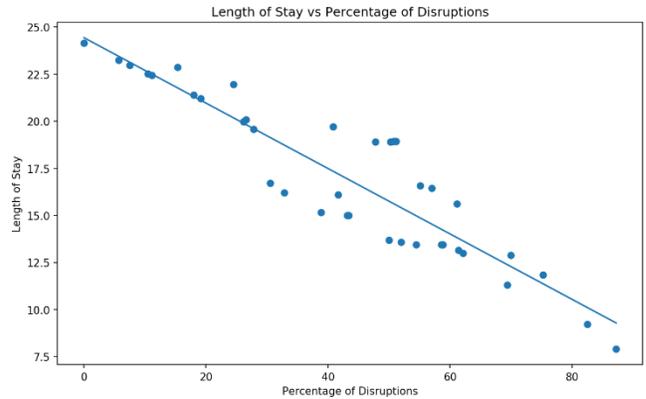

Fig. 5. Length of stay in ICU vs Percentage of disruptions.

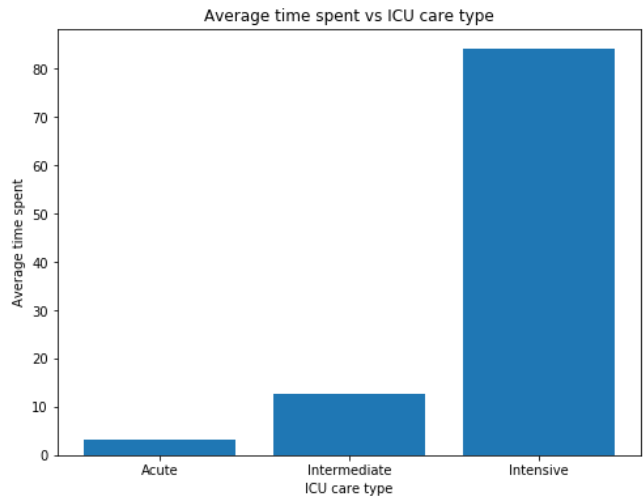

Fig. 6. Bar graph of Average Percentage of time spent by the patient for each care type namely Acute, Intermediate and Intensive

After applying LOESS smoothing to the data (Fig. 5), we can see that there is a negative correlation between length of stay and the percentage of disruptions for a patient. This could imply that patients likely undergo a greater number of interventions from the medical team in highly acute settings. Further using linear regression for the length of stay as the dependent variable vs the number of disruptions along with factors like quality of sleep, gender, age, and type of surgery, we found similar results of a negative correlation coefficient of -0.43. The P-value for correlation of the number of disruptions to the length of stay is 0.039, thus we can say that the number of disruptions does, in fact, influence the length of stay. The other factors have P-value > 0.05 and thus do not directly affect the length of stay but play a significant role when taken into consideration with disruptions.

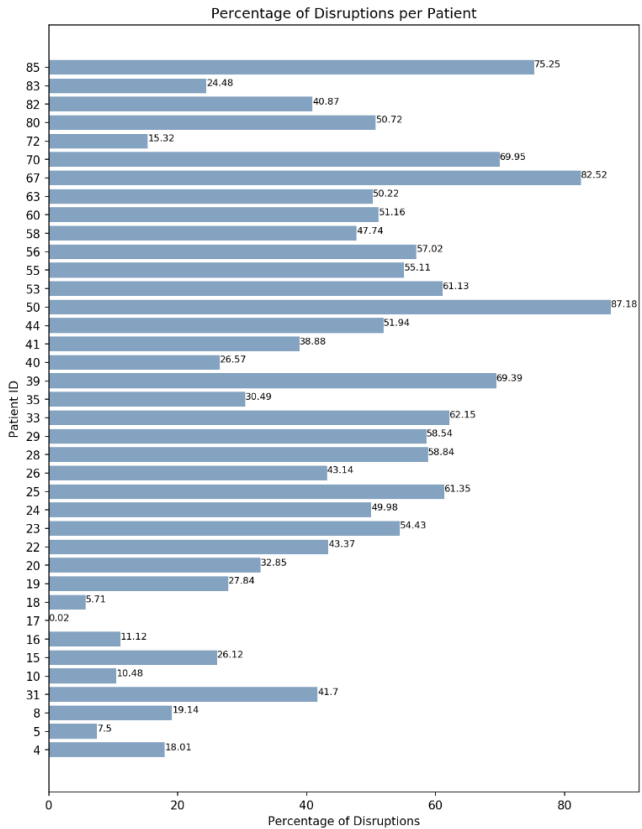

Fig. 7. Percentage of frames marked as disruptions for each patient.

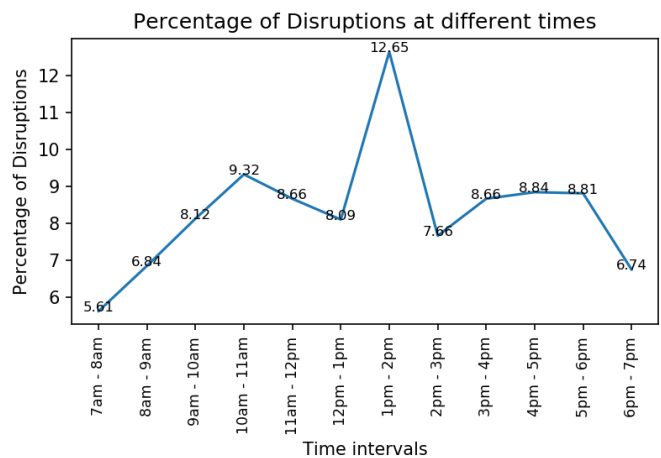

Fig. 8. Percentage of disruptions at different time intervals of the day.

Looking at Fig. 6, we can say that majority of the time spent in the ICU by a patient is of intensive type. Thus, the patients in ICU require additional care and special accommodations with respect to visitation hours and medical care disruptions should be made to ensure that the disruptions don't negatively affect the patient.

43% of the frames sampled in this study contained disruptions. The mean number of people in the room is 2.61 which means that there are normally 2-3 people in the room when the patient is being interrupted. From Fig. 7, it's clear that some patients undergo more disruptions than others. This might be due to higher visitations from relatives or due to the nature of treatment requiring more medical interventions.

From Fig. 8, the maximum disruptions occur between 1 pm – 2 pm which is likely the lunchtime while the least occur between 7 am – 8 am when the patient is usually asleep. As evening approaches, the number of disruptions goes down which is ideal as the patients tend to sleep and get rest.

## V. CONCLUSION

While this study provides a promising approach for automatically quantifying disruptions in the ICU, it also has several limitations. Our approach does not distinguish between daily visits and medical care and considers all as disruptions. Currently, we only analyze daytime videos to ensure sufficient lighting and good performance. In the future, we will perform finer-grained analysis of the scene, will recruit a larger sample, and will also analyze nighttime videos to compare disruptions throughout the 24 hours cycle. We also plan to perform the analysis in real-time without any need for recording the videos. This approach has the potential for improving sleep and rest quality in critically ill patients by providing a measure of rest disruptions in the ICU.

### ACKNOWLEDGMENT

This work is supported by NSF CAREER 1750192 (PR), and NIH/NIBIB 1R21EB027344 (PR). The Titan X Pascal partially used for this research was donated by the NVIDIA Corporation. We thank Mr. Matthew Rupert and Ms. Julie Cupka for their contributions during data collection. Dr. Azra Bihorac and Dr. Parisa Rashidi were supported by R01 GM110240 from the NIGMS, and by the National Institute of Biomedical Imaging and Bioengineering (grant R21EB027344-01). Dr. Parisa Rashidi was supported by CAREER award, NSF-IIS 1750192, from the National Science Foundation (NSF), Division of Information and Intelligent Systems (IIS).